\documentclass[aps,prc,twocolumn,superscriptaddress,amsmath,showpacs,amssymb]{revtex4-1}
\usepackage{bm,physics}

\usepackage[dvips,final]{graphicx}

\usepackage{ulem, color}
\usepackage{hyperref}
\bmdefine{\ba}{a}
\bmdefine{\bb}{b}
\bmdefine{\bx}{x}
\bmdefine{\by}{y}
\bmdefine{\bz}{z}
\bmdefine{\bn}{n}
\bmdefine{\bp}{p}

\newcommand{\BM}{\begin{pmatrix}}
\newcommand{\EM}{\end{pmatrix}}

\renewcommand{\d}{\dagger}

\newcommand{\Lc}{\mathcal{L}}
\newcommand{\Mc}{\mathcal{M}}

\newcommand{\hphi}{\hat\varphi}
\newcommand{\hpsi}{\hat\psi}

\newcommand{\intx}{\int\! d^3x\;}
\newcommand{\intxd}{\int\! d^3x'\;}
\newcommand{\intxxd}{\int\! d^3x\,d^3x'\;}

\newcommand{\ex}{\mathrm{ex}}

\begin{document}
\title {%Evidence of
%for
Bose--Einstein condensation of  dilute  alpha clusters
   above four $\alpha$ threshold\\
in  $^{16}$O in field theoretical superfluid cluster model
}
\author{J.~Takahashi}
\email{takahashi.j@aoni.waseda.jp}
\affiliation{Department of Electronic and Physical Systems,
Waseda University, Tokyo 169-8555, Japan}

\author{Y.~Yamanaka}
\email{yamanaka@waseda.jp}
\affiliation{Department of Electronic and Physical Systems,
Waseda University, Tokyo 169-8555, Japan}

\author{S.~Ohkubo}
\email{shigeo.ohkubo@rcnp.osaka-u.ac.jp}
\affiliation{Research Center for Nuclear Physics,
Osaka University, Ibaraki,
Osaka 567-0047, Japan}

\date{\today}

\begin{abstract}
Observed well-developed $\alpha$ cluster states   in $^{16}$O located  above the four $\alpha$ threshold are investigated from the viewpoint of Bose-Einstein condensation of $\alpha$ clusters by using a field-theoretical superfluid cluster model  in which  the order parameter is defined.  The experimental energy levels are  reproduced well for the first time by calculation.
In particular, the observed 16.7 MeV $0_7^+$ and 18.8 MeV $0_8^+$ states  with low-excitation energies from the threshold   are found  to be understood as a manifestation of  the states of the Nambu-Goldstone zero-mode operators, associated with the spontaneous symmetry breaking of the global phase, which is caused by the Bose-Einstein condensation of the vacuum 15.1 MeV $0^+_6$ state with a dilute well-developed $\alpha$ cluster structure just above the threshold.
This gives evidence  of the existence of the Bose-Einstein condensate of $\alpha$ clusters in $^{16}$O.
 It is found that the   emergence of the energy level structure with a well-developed $\alpha$ cluster structure above the threshold   is robust, almost independently of the condensation rate of  $\alpha$ clusters under significant  condensation rate.
The finding of the mechanism why  the  level structure that is similar to  $^{12}$C emerges   above the four $\alpha$ threshold in $^{16}$O  reinforces  the concept of BEC of $\alpha$ clusters    in addition to $^{12}$C.
 \end{abstract}

\maketitle

\par
In this paper, we present evidence of Bose-Einstein condensation (BEC) of $\alpha$ clusters in $^{16}$O.
 The idea of  cluster structure of nuclei was originally proposed in Refs.~\cite{Wefelmeier1937,Wheeler1937}  as the first nuclear structure model and
since the proposal of the Ikeda diagram   in $^{8}$Be$-$$^{24}$Mg \cite{Ikeda1968} and   $^{8}$Be$-$$^{32}$S \cite{Horiuchi1972}, the  last   century  witnessed  a fruitful existence of cluster structure in light nuclei  \cite{Suppl1972,Wildermuth1977,Suppl1980}. The  cluster structure has been shown to persist  also in medium-weight and heavy mass region \cite{Suppl1998}, where the spin-orbit force
 becomes important. The diagram was  extended to the  $fp$-shell region, $^{44}$Ti region in Ref.~\cite{Ohkubo1989} and  $^{60}$Zn region in Ref.~\cite{Ohkubo1998}.
 The  discoveries of the excitations
 of  the inter-cluster  relative (vibrational) motion  (the higher nodal state)
  typically in $^{20}$Ne \cite{Fujiwara1980} and $^{44}$Ti \cite{Michel1998}, which  is not expected from  the mean field  picture of nuclei  \cite{Bohr1969B,Ring1980}, showed a {\it diversity} of collective motion caused by clustering.

\par
As a new {\it diversity} of collective motion,
BEC of $\alpha$ clusters has  been attracting
much attention
 in the  $N\alpha$ systems in $4N$ nuclei, such as  three $\alpha$ clusters in $^{12}$C and   four $\alpha$ clusters in $^{16}$O,
which  are the highest order clustering
 in the Ikeda diagram~\cite{Ikeda1968,Horiuchi1972}. This is a new  collective motion caused by the spontaneous symmetry breaking (SSB) of the global phase in the {\it gauge space}  and different from the  collective motions  in {\it configuration space} such as rotation and vibration with a geometrical intrinsic  structure, which
  are also widely seen in the mean field  picture
   \cite{Bohr1969B,Ring1980}.

\par
In fact, three $\alpha$ clusters in $^{12}$C($0_2^+$, 7.654 MeV), the Hoyle state,  has been shown to be in a gas phase
 by Uegaki {\it et al.} \cite{Uegaki1977} in contrast to the preceding picture of a rigid three $\alpha$ chain structure \cite{Morinaga1956}.
    Matsumura {\it et al.} \cite{Matsumura2004} showed  that  about 70\% of the $\alpha$ clusters in the Hoyle state  are sitting in the 0s state. Many theoretical and experimental  studies were devoted to investigate whether  the Hoyle state is  a   Bose-Einstein condensate \cite{Matsumura2004,Tohsaki2001,Kurokawa2004,Matsumura2004,Yamada2004,Ohtsubo2013,Funaki2015,%
    Funaki2016,Kanada2007,Chernykh2007,Roth2011,Epelbaum2012,Dreyfuss2013,Nakamura2016,Katsuragi2018}.

 \par
  The most  intriguing
   nucleus beyond $^{12}$C  is $^{16}$O, for which  systematic experimental data have been accumulated \cite{Chevallier1967,Freer1995,Freer2004,Freer2005,Itoh2014,Curtis2016}.
Despite  theoretical studies \cite{Tohsaki2001,Funaki2008C,Ohkubo2010,Funaki2018} whether  the observed energy {\it level structure} of the  $\alpha$ cluster states     just above the four $\alpha$ threshold in $^{16}$O is due to BEC of $\alpha$ clusters  has not been   confirmed.
Full microscopic  four $\alpha$ cluster model calculations  and   {\it ab initio} calculations to understand the $\alpha$ cluster level structure
are presently formidably difficult for $^{16}$O.
  Also we note that  whatever
   the calculated   rate of $\alpha$ clusters sitting in the 0s state may be, any theory without  the order parameter
   is unable  to conclude whether  the system  is in the {\it Nambu-Goldstone (NG) phase} of BEC or   in the normal {\it Wigner phase}.

\par
   We have proposed a field theoretical superfluid cluster model in which the order parameter is defined \cite{Katsuragi2018,Nakamura2016} and showed in  $^{12}$C that the emergence of    the peculiar energy levels with a well-developed gas-like  $\alpha$ cluster structure built on the Hoyle state
    is  a manifestation of   the  NG phase  caused by  the BEC  of the vacuum Hoyle state.

\par
 The purpose of this paper is to show  for the first time that the observed peculiar  $\alpha$ cluster level structure  just above the four $\alpha$ threshold  in $^{16}$O can be understood     in a  superfluid $\alpha$ cluster model
   and that in particular the emergence   of the  ground and excited
states of the NG operators is  a manifestation of  the evidence  for the    BEC  of  $\alpha$ clusters.

\par
From the theoretical side, Tohsaki {\it et al.}  \cite{Tohsaki2001} considered  that  the $0_5^+$  (14.0 MeV) state   in $^{16}$O, located 0.44 MeV only just below the four $\alpha$ threshold energy, is a BEC state of  $\alpha$ clusters.
On the other hand, Funaki {\it et al.} \cite{Funaki2008C} discussed, using a four $\alpha$ semi-microscopic orthogonality condition model (OCM),
 that  the $0_6^+$ (15.1 MeV) state   located just above the four $\alpha$ threshold \cite{ENSDF} is a  condensate of  four $\alpha$ clusters.  Ohkubo {\it et al.} \cite{Ohkubo2010}
   suggested,  in the unified description of  nuclear rainbows in $\alpha$+$^{12}$C scattering and $\alpha$ cluster structures in the bound and quasi-bound  energy region of $^{16}$O,
that  the  observed $0_6^+$ (15.1 MeV) state could be    a superfluid state of  four $\alpha$ clusters.
As for the excited states above the $0_6^+$ (15.1 MeV) state,  i.e.,  $2^+$  (16.95 MeV), $2^+$ (17.15 MeV), $4^+$ (18.05 MeV) and $6^+$ (19.35 MeV) states, observed by Chevalier {\it et al.}  \cite{Chevallier1967}, which had been considered  to have a four $\alpha$ linear chain structure \cite{Chevallier1967,Horiuchi1972,Suzuki1972},
  Ohkubo {\it et al.} \cite{Ohkubo2010} showed that   they can be understood  to have the local condensate  $\alpha$+$^{12}$C ($0_2^+$) structure.
As for the four $\alpha$  linear chain in $^{16}$O,
 Ichikawa {\it et al.} \cite{Ichikawa2011}  and others \cite{Horiuchi2017,Inakura2018} showed  that  the   excitation energy of the band head state  is much higher, above 30 MeV.
 A search to observe such a high lying  state has been attempted \cite{Curtis2013}.

\par
On the other hand, from the experimental side,
very important  developments in searching for BEC of  $\alpha$ clusters in $^{16}$O  were recently reported by Itoh {\it et al.} \cite{Itoh2014}, who  observed  two broad resonant  $0^+$ states, i.e., $0_7^+$ at
16.7 MeV with the $\alpha$+$^{12}$C($0_2^+$) structure and $0_8^+$ at 18.8 MeV with the
 $^8$Be+$^8$Be structure just above the  $0_6^+$ (15.1 MeV) state.
 We note that this energy 16.7 MeV of   $0_7^+$ exactly agrees within the width with 16.6 MeV predicted by Ohkubo {\it et al.} \cite{Ohkubo2010}  from the viewpoint of   BEC and  superfluidity  of $^{16}$O.
  In Ref.~\cite{Ohkubo2013}, Ohkubo  suggested that the  $0_7^+$ state   (16.7 MeV) may be a  NG state due to the spontaneous symmetry breaking of the global phase caused by the BEC of the  $0_6^+$ (15.1 MeV) state.
 In addition to the experimental data by Chevallier {\it et al.} \cite{Chevallier1967} and Freer {\it et al.} \cite{Freer1995,Freer2004,Freer2005},
 Curtis {\it et al.}  \cite{Curtis2016} very recently  observed in the $^{13}$C($^4$He,\,4$\alpha$)n breakup reaction the $2^+$ (17.3 MeV),  $4^+$ (18.0 MeV),  $2^+$ or $4^+$ (19.4 MeV), and  $4^+$ or $6^+$  (21.0 MeV)  states, which decay into the $^8$Be+$^8$Be channel. These observations seem to be consistent with the previous results by Freer {\it et al.} \cite{Freer1995},
 the  $2^+$  (17.1 MeV),  $2^+$ (17.5 MeV),  $4^+$ (19.5 MeV) and $6^+$ (21.4 MeV) states in the  $^{12}$C($^{16}$O,\,4$\alpha$)$^{12}$C reaction, and by Freer {\it et al.} \cite{Freer2004},
  the $6^+$ states at 20.0 and 21.2 MeV  in the $^8$Be+$^8$Be decay channel of  $^{16}$O.
In the present study, we focus  on the  well developed four $\alpha$ states above 
the four $\alpha$ threshold (14.44 MeV), because
almost all the energy levels  and the  electric transition probabilities below about
 13 MeV in excitation energy are known to be explained well in the $\alpha$+$^{12}$C cluster model by Suzuki \cite{Suzuki1976}.

\par
We study these $\alpha$ cluster states from the viewpoint of BEC of $\alpha$ clusters,
using the field-theoretical superfluid  $\alpha$  cluster model
~\cite{Nakamura2016,Katsuragi2018}.
We  briefly recapitulate  the formulation.
The
 model Hamiltonian for a bosonic field $\hpsi(x)$
$(x=(\bx,t))$ representing
the $\alpha$ cluster is given as follows:
\begin{align}
&\hat{H}=\intx \hpsi^\d(x) \left(-\frac{\nabla^2}{2m}+
V_\ex(\bx)- \mu \right) \hpsi(x) \notag\\
&\,\,+\frac12 \intxxd \hpsi^\d(x)
\hpsi^\d(x') U(|\bx-\bx'|) \hpsi(x') \hpsi(x) \,.
\label{Hamiltonian}
\end{align}
Here, the potential $V_\ex$ is a mean field potential introduced phenomenologically
to trap the $\alpha$ clusters inside the nucleus, and is taken to have a harmonic
form,
$
 V_\ex(r)= m \Omega^2 r^2/2\,,
$
and the {\it residual} $\alpha$--$\alpha$ interaction
is given by
the Ali--Bodmer type two-range Gaussian potential \cite{Ali1966},
$
U (|\bm x -\bm x'|) = V_r e^{-\mu_r^2 |\bm x -\bm x'|^2}
- V_a e^{-\mu_a^2 |\bm x -\bm x'|^2}\,.
$
We  need no phenomenological three-body and four-body
$\alpha$--$\alpha$ interactions \cite{Funaki2008C}.
We set $\hbar=c=1$ throughout this paper.

\par
When BEC of $\alpha$ clusters occurs, i.e.
the global phase symmetry of $\hpsi$ is spontaneously broken, we decompose $\hpsi$
as $\hpsi(x)=\xi(r)+\hphi(x)$, where the c-number $\xi(r)=\bra{0} \hat\psi(x)
\ket{0}$ is an order parameter and is assumed to be real and isotropic.
To obtain the excitation spectrum, we need to solve
three coupled equations, which are the Gross--Pitaevskii (GP) equation, Bogoliubov-de Gennes (BdG) equations, and zero-mode equation \cite{Nakamura2016, Katsuragi2018}.
The GP equation that determines the order parameter
 is given by
\begin{equation}\label{eq:GP}
\left\{ -\frac{\nabla^2}{2m}+V_\ex(r) -\mu + V_H(r)
\right\} \xi(r) = 0 \,,
\end{equation}
where
$
    V_H(r) = \intxd U(|\bx-\bx'|)\xi^2(r')\,.
$  The order parameter $\xi$ is normalized
with the  condensed particle number $N_0$ as
\begin{align}
\intx |\xi(r)|^2 = N_0\,.
\end{align}
The BdG equations that describe the collective oscillations on the condensate
are given by
\begin{align}
  \intxd
  \left(\begin{array}{cc}
        \Lc & \Mc \\
        -\Mc^* & -\Lc^*
      \end{array}\right)
  \left(\begin{array}{c}
    u_{\bn} \\
    v_{\bn}
   \end{array}\right)
  = \omega_\bn
    \left(\begin{array}{c}
        u_{\bn} \\
        v_{\bn}
\end{array}  \right),
\end{align}
where
$
  \Mc(\bx, \bx')
   = U(|\bx-\bx'|) \xi(r) \xi(r'),\,
  \Lc(\bx, \bx')
   = \delta(\bx-\bx')
     \left\{ -\frac{\nabla^2}{2m}+V_\ex(r) -\mu + V_H(r)\right\}
     + \Mc(\bx, \bx')\,.
$
The index $\bm{n}=(n,\, \ell,\, m)$ stands for the main, azimuthal and
magnetic quantum numbers. The eigenvalue $\omega_\bn$ is the excitation energy
of the Bogoliubov mode. For isotropic $\xi$, the BdG eigenfunctions can be take to have separable forms,
$u_{\bm{n}}(\bm{x}) = \mathcal{U}_{n\ell}(r) Y_{\ell m}(\theta, \phi), \,
 v_{\bm{n}}(\bm{x}) = \mathcal{V}_{n\ell}(r) Y_{\ell m}(\theta, \phi).
$
We necessarily have an eigenfunction belonging
to zero eigenvalue, explicitly $(\xi(r), -\xi(r))^t$, and its adjoint function
$(\eta(r),\eta(r))^t$ is obtained as
\begin{equation}
\eta(r)= \frac{\partial}{\partial N_0}\xi(r)\,.
\end{equation}
The field operator is expanded as
\begin{align}
\hphi(x)&=-i{\hat Q}(t)\xi(r)+{\hat P}\eta(r) \notag \\
&\,\, +\sum_{\bn} \left\{{\hat a}u_\bn(\bx)
+{\hat a}^\dagger v^\ast_\bn(\bx) \right\}\,,
\end{align}
with the commutation relations $[{\hat Q}\,,\,{\hat P}]=i$ and
$[{\hat a}_{\bn}\,,\,{\hat a}^\dagger_{\bn'}]=
\delta_{\bn \bn'}$\,. The operator ${\hat a}_\bn$ is an annihilation operator
of the Bogoliubov mode, and the pair of canonical operators ${\hat Q}$ and ${\hat P}$
originate from the SSB of the global phase and are called the NG or zero-mode operators.

The treatment of the zero-mode operators is a chief feature of our approach.
The naive choice of the unperturbed bilinear Hamiltonian with respect
to ${\hat Q}$ and ${\hat P}$ fails due to their large quantum fluctuations.
Instead, we gather all the terms consisting only of ${\hat Q}$ and ${\hat P}$
in the total Hamiltonian to construct the unperturbed nonlinear Hamiltonian
of ${\hat Q}$ and ${\hat P}$, denoted by $H_u^{QP}$\,.
The coefficients in $\hat{H}_u^{QP}$ are $I=\frac{\partial \mu}{\partial N_0}$
and the integrations involving $\xi\,,\,\eta\,$ and $U$, whose explicit forms are
seen in the Ref.~\cite{Katsuragi2018}. We set up the eigenequation
for $\hat{H}_u^{QP}$, called
the zero--mode equation,
\begin{align} \label{eq:HuQPeigen}
\hat H_u^{QP} \ket{\Psi_\nu} = E_\nu \ket{\Psi_\nu}\qquad
(\nu=0,1,\cdots)\,.
\end{align}
This equation is similar to a one-dimensional
Schr\"odinger equation for a bound problem.

\par
The total unperturbed Hamiltonian ${\hat H}_u$ is ${\hat H}_u=\hat H_u^{QP}
+\sum_{\bn} \omega_\bn {\hat a}_\bn^\dagger{\hat a}_\bn$.
The ground state energy is set to zero, $E_0=0$.
The states that we consider are $\ket{\Psi_\nu}\ket{0}_{\rm ex}$ with
energy $E_\nu$, called the zero-mode state, and
$\ket{\Psi_0}{\hat a}^\dagger_\bn
\ket{0}_{\rm ex}$ with energy $\omega_\bn$, called the BdG state,
where ${\hat a}_\bn \ket{0}_{\rm ex}=0$.

\par
As in Refs.~\cite{Nakamura2016,Katsuragi2018}, we
 adjust the     strength  parameter $V_r$ of the short range  repulsive potential of the Ali-Bodmer potential (set $d0$) \cite{Ali1966}
 as a {\it residual} interaction in the mean field  potential $V_{\rm ex}$,
  because it stabilizes the condensate under the trapping potential and is the most sensitive in our analysis.
The repulsive   Coulomb potential affects numerical results little as in the case of $^{12}$C \cite{Nakamura2016,Katsuragi2018}
in the presence of the mean field  potential $V_{\rm ex}$, and is suppressed  in the present work.
The chemical potential
is fixed by the input $N_0$.

\par
We identify  $0_6^+$ (15.10 MeV) state, considered to be a Hoyle analog-state \cite{Funaki2008C,Ohkubo2010}, as the vacuum $\ket{\Psi_0}\ket{0}_{\rm ex}$.  In fact,
according to Ref.~\cite{Funaki2008C},
   61\% of  the $\alpha$ clusters in this state are sitting in the 0s state, while the  rate is less than 25\% for all the  other  $0^+$ states below, including  $0_5^+$  (14.0 MeV).
The two parameters,
 $\Omega=1.57$ MeV$/ \hbar$ and $ V_r$=514 MeV,
which are  determined by inputting $N_0=0.61\times4$ and
$E_1=$16.7-15.1=1.6 MeV when
the  experimental $0_7^+$  (16.7 MeV) state is identified as
$\ket{\Psi_1}\ket{0}_{\rm ex}$ ($0_{\nu=1}^+$(ZM)) and also
by minimizing the mean square errors between the experimental
(centroid energies for the fragmented  $2^+$, $4^+$ and $6^+$ states)
 and calculated energies of the states above it, and
to reproduce well the excitation energy  of the  experimental $0_7^+$  (16.7 MeV) state from the vacuum  $0_6^+$ (15.10 MeV), i.e., $E_1=$16.7-15.1=1.6 MeV under  the condensation rate 61\%, are used in the calculations below.

%Fig1  experimental energy levels of 16O  above the 4 alpha threshold
\begin{figure}[t]
\begin{center}
\includegraphics[width=8.5cm]{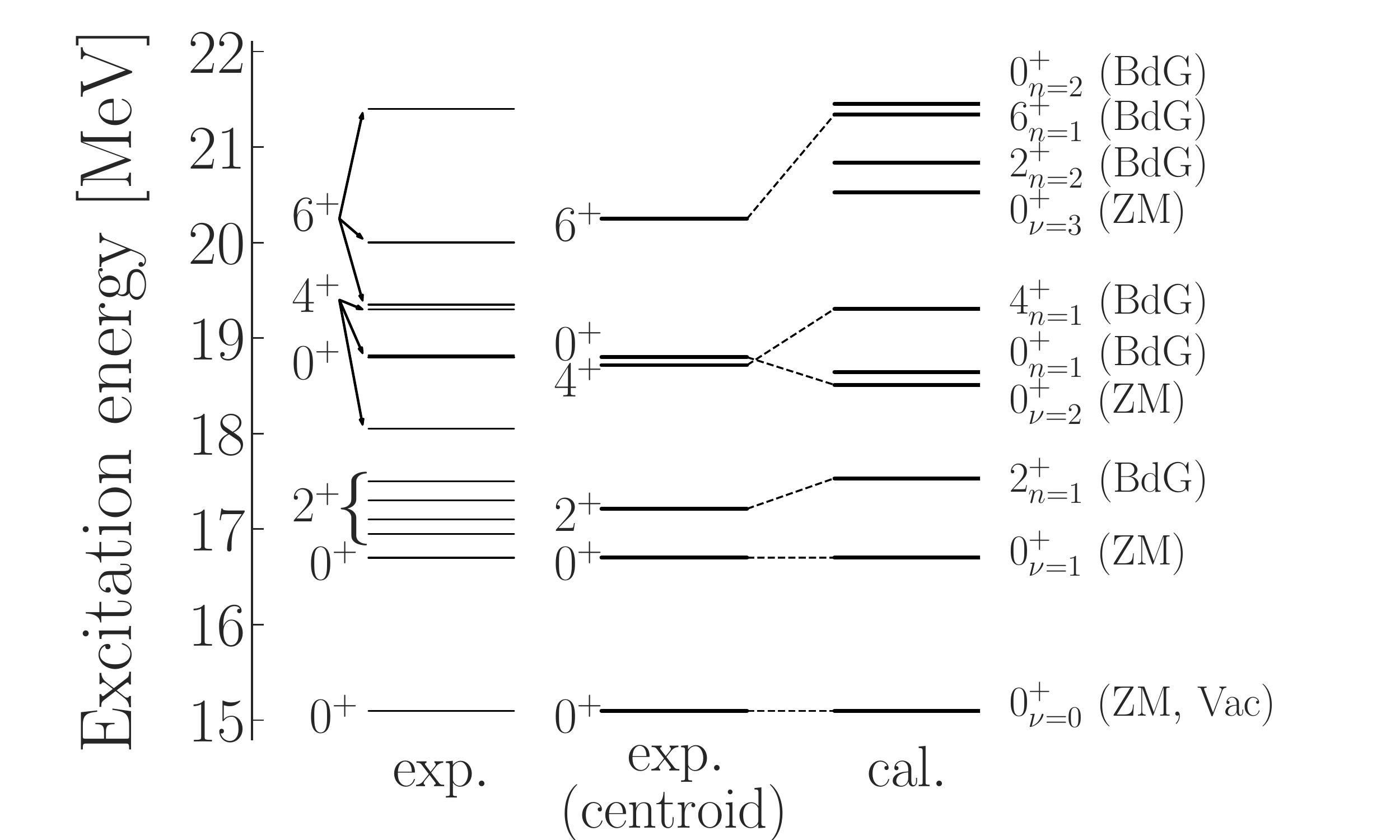}
\caption{Energy levels in $^{16}$O calculated
with 61\% condensation rate on the right
are compared with the experimentally  spin assigned  $\alpha$ cluster states, $0^+$ (16.7 and 18.8 MeV)  from Refs.\cite{Itoh2014,ENSDF},  $2^+$ (16.95, 17.1,  17.3 and 17.5 MeV),
 $4^+$ (18.05, 18.8  and 19.30 MeV) and $6^+$  (19.35, 20.0 and 21.4 MeV) from Refs.\cite{Chevallier1967,Freer1995,Freer2004,Curtis2016} on the left.
The centroid red energies of the respective
 $2^+$, $4^+$ and $6^+$ states are shown in the middle.
% they may be fragmented from and also may coincide with other states considering  the experimental resolution 0.35 MeV \cite{Freer1995},  0.2 MeV \cite{Curtis2016}  and $<0.2$ keV\cite{Freer2004}  and to void complexity only the centroid energy levels are drawn in Fig.~1.
 % $2^+$ (17.1 MeV and 17.5 MeV) in Ref.~\cite{Freer1995}
 %  is displayed.}
}
\label{fig:Energy_levels}
\end{center}
\end{figure}

\par
In Fig.~\ref{fig:Energy_levels}, the energy levels calculated  using the obtained
parameters ($\Omega$, $V_r$)
are displayed.
We notice that the calculated
 second zero-mode state $0_{\nu=2}^+$(ZM)   appears at low excitation energy from the vacuum $0_6^+$(15.10 MeV) and agrees  well  with the  broad $0_8^+$(18.8 MeV) state  observed  by Itoh {\it et al.} \cite{Itoh2014}.  Our calculation locates also the $0_{n=1}^+$(BdG)  near the zero-mode  $0_{\nu=2}^+$ (ZM) state, which seems to be within the broad width of the experimental $0_8^+$(18.8 MeV) state.  From the excitation function of Fig.~4(b) in  Ref.~\cite{Itoh2014}, it is difficult to know whether the broad  $0_8^+$(18.8 MeV) state is fragmented.   The calculated BdG excited states $2_{n=1}^+$(BdG), $4_{n=1}^+$(BdG) and $6_{n=1}^+$(BdG) states  agree with the
 centroids of the
  experimental energy levels observed  in
  \cite{Chevallier1967,Freer1995,Freer2004,Curtis2016}.
  Our calculation predicts a $8_{n=1}^+$ (BdG) at
  {23.7 MeV} and a   $10_{n=1}^+$(BdG)   at
  { 26.4 MeV}.
  The calculated $8^+$   state may correspond
 to the  $8^+$ state at  23.6 MeV observed
   in Ref.~\cite{Freer2004}.

\par
The logical  structure that the  energy levels of  the two $0^+$  zero-mode states appear at  low energies  from the vacuum
 followed by  the $2^+$(BdG) and $4^+$(BdG) states in agreement with
 experiment  is the
 same as that of the excited $\alpha$ cluster states  above the Hoyle state in $^{12}$C.
  The level structure is a manifestation of
   the emergence of the  zero-mode and BdG states due to BEC of the $\alpha$ clusters.
     The appearance of the zero-mode states  confirms the conjecture  by Ohkubo in Ref.~\cite{Ohkubo2013}.

 \par       
   The reason   why the dilute well-developed four $\alpha$ cluster condensate states
 are fragmented was investigated by one of the present author (SO) and  Hirabayashi in the coupled channel calculations that include the non-dilute  $\alpha$+$^{12}$C(gs, $2_1^+(4.44$ MeV), $3^-(9.65$ MeV)) cluster structure  and the dilute $\alpha$+$^{12}$C($0_2^+(7.65 $MeV), $2_2^+(10.3$ MeV) )  cluster structure \cite{Ohkubo2010}. It was shown that  the broad dilute  $\alpha$+$^{12}$C (Hoyle state) structure  is fragmented,  typically for the 4$^+$ state, into two  states  separated about 1.5 MeV by the coupling.  This is easily understood qualitatively by noting  that the unperturbed energy of the $\alpha$+$^{12}$C($2_1^+$, 4.44 MeV) cluster structure with the relative angular momentum $L=4$, which generates [$2_1^+$$\otimes$ $L=4$]$_{J=2^+,4^+,6^+}$, located at 15.5 MeV in the relevant energy region.  The appearance of the three  fragmented   states for each spin   $2^+$, $4^+$, $6^+$  by the coupling can be understood qualitatively by noting the additional presence of the   $\alpha$+$^{12}$C($4^+$, 14.08 MeV) cluster structure with  $L=2$,  which    generates
% [$4^+$$\otimes$ $L=2$]$_
$ {J=2^+,4^+,6^+}$ at the  unperturbed energy  21.0 MeV in the relevant energy region.
 The {\it fragmentation of BEC states} of the $\alpha$ clusters was not observed in the BEC of the three $\alpha$ clusters in $^{12}$C, in which no other excited state exists except the  $2_1^+$ below the three $\alpha$ threshold. The interference between the BEC state  and  non-BEC impurity  states is a new feature that has never been known and is  expected to appear in other nuclei.
 In our model the fragmentation of the BEC can be treated  by introducing  a scalar field corresponding to the non-dilute $\alpha$+$^{12}$C configuration, which is  beyond the scope of the  present work and is a future challenge.

 %Fig.2 order parameter
\begin{figure}[t]
\begin{center}
\includegraphics[width=8.6cm]{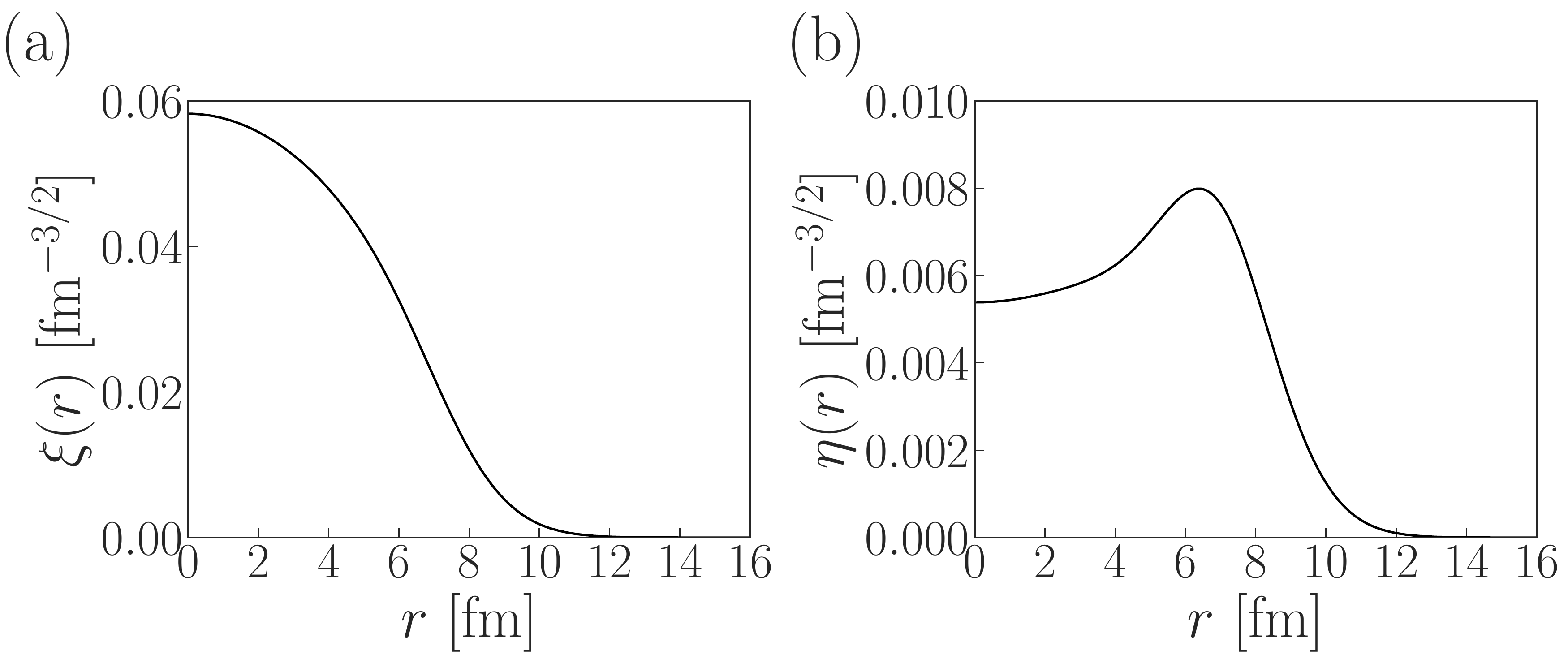}
\caption{ (a) the calculated  order parameter $\xi(r)$ and (b) its adjoint eigenfunction $\eta(r)$ with 61$\%$ condensation rate.
}
\label{fig:xi}
\end{center}
\end{figure}

\par
In Fig.~\ref{fig:xi}, the density distribution of the order parameter $\xi(r)$, which is related to the superfluidity density $\rho(r)$=$|\xi(r)|^2/N_0$, is displayed.
The rms radius of the Hoyle-analog $0_6^+$  (15.1 MeV)  state ${\bar r}$ is calculated  to be 5.22 fm as ${\bar r}^2= \intx\, r^2|\xi(r)|^2/ N_0$.
Although the experimental value is not available, this   large size  compared with the experimental  2.71 fm  of the ground state of $^{16}$O  supports that the Hoyle-analog state is a dilute $\alpha$ cluster state.   This value is also  consistent with  5 fm  in Ref.~\cite{Funaki2008C}.
In Fig.~\ref{fig:xi}~(a) the superfluid density is the largest in the center of the condensate and decreases gradually toward the surface.  We note that the density extends  considerably beyond the rms radius  5.2 fm up to 10 fm.
 In Fig.~\ref{fig:xi}~(b) the calculated  $\eta$, which is the derivative of $\xi$ with respect to the number of $\alpha$ clusters, represents the number fluctuation  of the superfluid condensate.
 We note that the fluctuation is the largest not in the central region but
at around 6 fm in the surface region.
This behavior is similar to the $^{12}$C case in Ref.~\cite{Katsuragi2018}. The number fluctuation is visible even beyond 10 fm up to 12 fm where the superfluid density is very low. This feature does not change if the condensation rate is significantly increased   (for example 100\%)  or  decreased from the present 61\%.

 %Fig.3 BdG wave functions
\begin{figure}[t]
\begin{center}
\includegraphics[width=8.6cm]{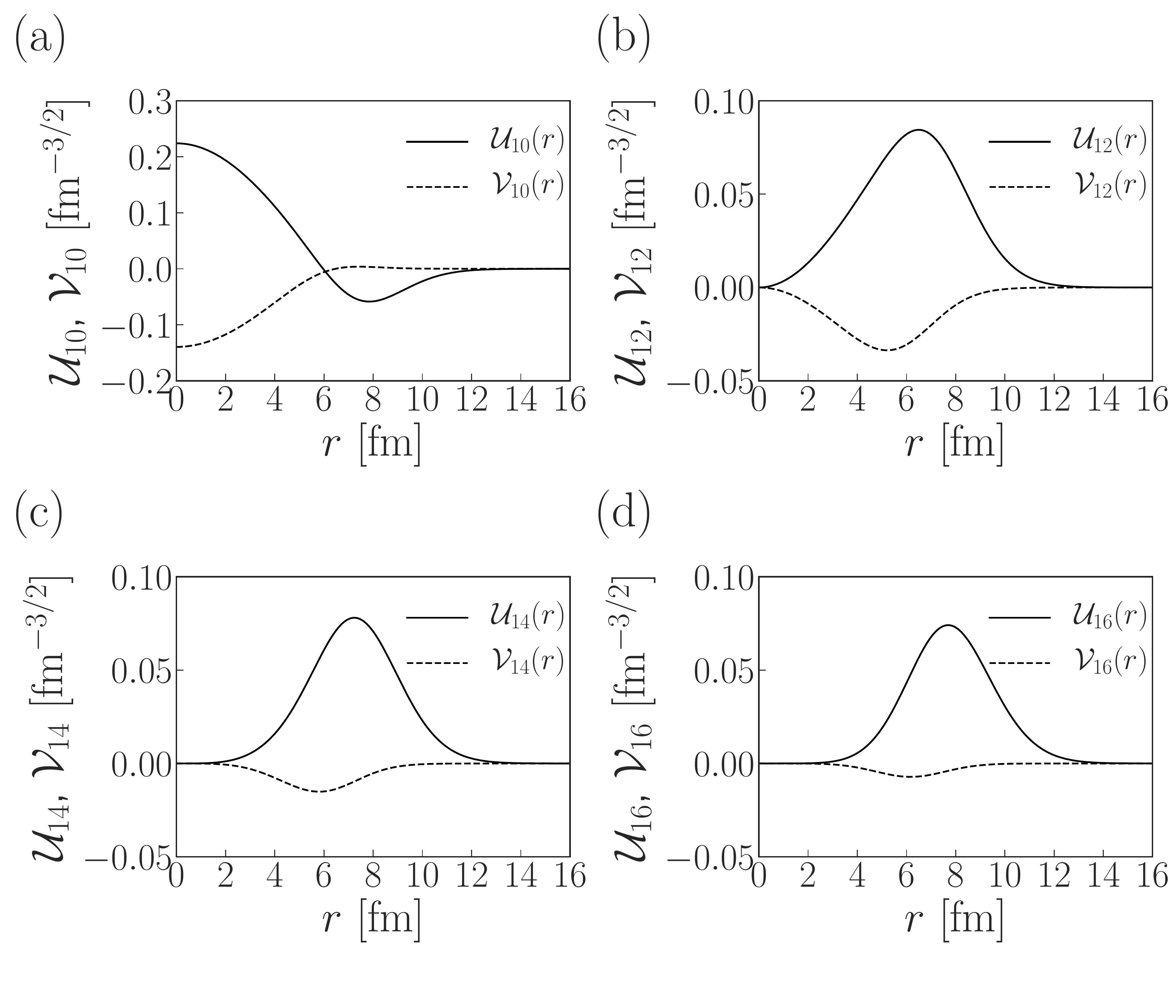}
\caption{The calculated  BdG wave functions $\mathcal{U}_{1\ell}(r)$ and $\mathcal{V}_{1\ell}(r)$ for the states  (a) $0^+$,
 (b) $2^+$,
   (c) $4^+$
     and (d) $6^+$
        with 61$\%$  condensation rate.
}
\label{fig:Wave_BdG}
\end{center}
\end{figure}

\par
 In Fig.~\ref{fig:Wave_BdG}, the wave functions $\mathcal{U}_{1\ell}(r)$ and $\mathcal{V}_{1\ell}(r)$ of the BdG states  ($n=1$) with $\ell$=0, 2, 4 and 6 are displayed.  We note that $|\mathcal{V}_{1\ell}(r)|$ is much smaller than $|\mathcal{U}_{1\ell}(r)|$, which becomes more notable as $\ell$ increases.
The peak of $\mathcal{U}_{1\ell}(r)$ for $\ell \neq 0$ is in the surface region because of the repulsive force of the central condensate and moves outward with increasing $\ell$ due to the centrifugal force.
For $\ell=0$, the orthogonality of $\mathcal{U}_{10}(r)$ and $\mathcal{V}_{10}(r)$ to the nodeless $\xi$  and $\eta$ makes the node at around 6 fm and the energy level is raised above the $\ell=2$ state. The amplitude of $\mathcal{V}_{1 0}(r)$ is not small even in the central region.
These features are similar to the BEC states built on the Hoyle state in $^{12}$C \cite{Nakamura2016,Katsuragi2018}.

%Fig.4  zero-mode wave functions
% Fig zero-mode wf
\begin{figure}[t]
\begin{center}
\includegraphics[width=7.8cm]{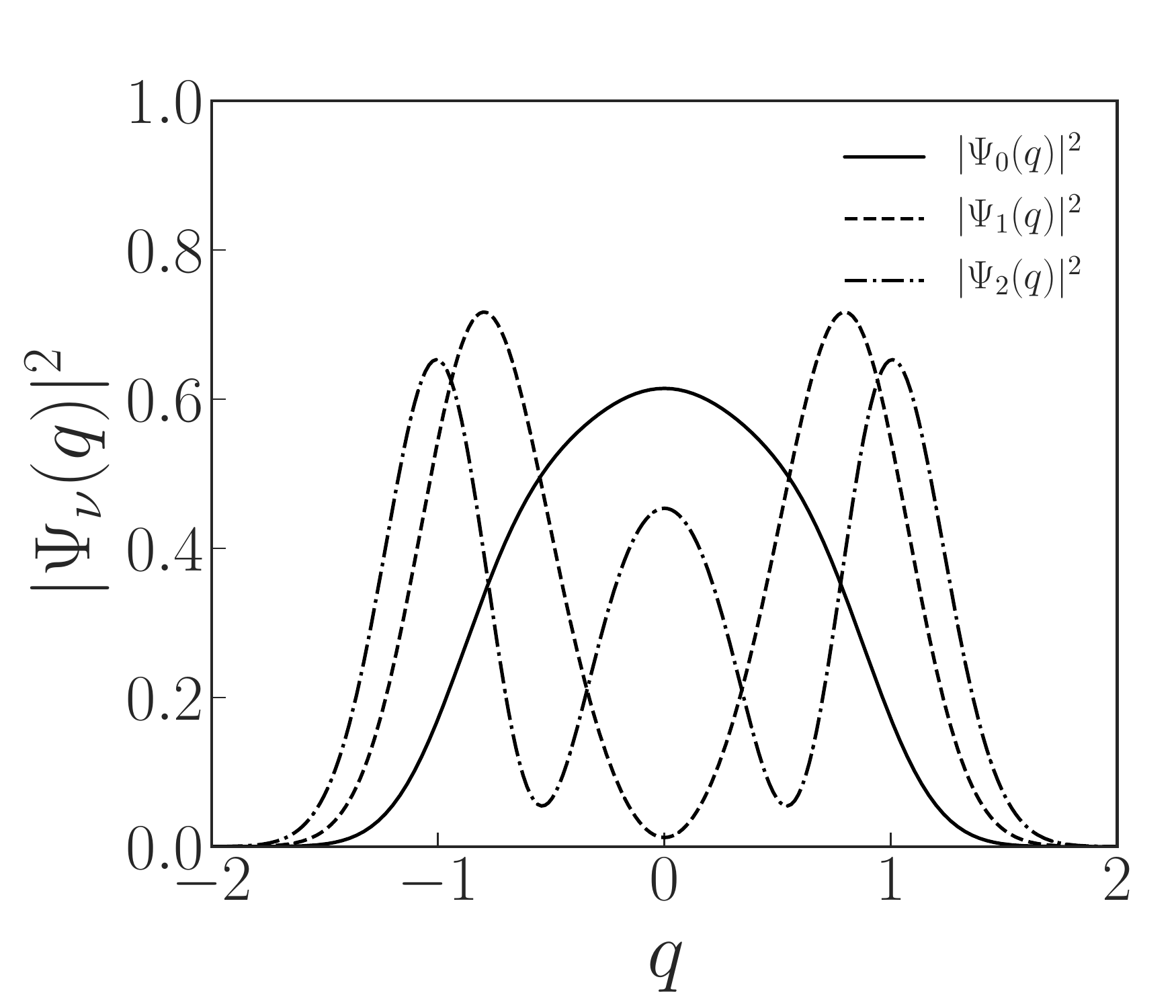}
\caption{The calculated  Nambu-Goldstone zero-mode wave functions
(a) $|\Psi_{0}(q)|^2$ ($0_{\nu=0}^+$(ZM,Vac)),
(b) $|\Psi_{1}(q)|^2$ ($0_{\nu=1}^+$(ZM)),
(c) $|\Psi_{2}(q)|^2$ ($0_{\nu=2}^+$(ZM))
 with 61$\%$  condensation rate.
}
\label{fig:WFs_zeromode}
\end{center}
\end{figure}

\par
In Fig.~\ref{fig:WFs_zeromode}, the squares of the  calculated  zero-mode wave functions  $|\Psi_\nu|^2$  for the first three ($\nu$=0, 1 and 2) states   are displayed.
The wave function of the $\nu$=0 zero-mode state
or the vacuum has no node in the $q$-space, as in  Fig.~\ref{fig:WFs_zeromode}~(a).
It is important to emphasize that in the present theory we obtain a series of $0^+$ states, which are the zero-mode excited states with $\nu\neq 0$.
These excited states are realized by increasing the number of the nodes of the wave function in the $q$-space, as seen in  Fig.~\ref{fig:WFs_zeromode}~(b) and (c).  The first excited  state with one node ($\nu$=1) in  Fig.~\ref{fig:WFs_zeromode}~(b) and the second excited state ($\nu$=2) with two nodes  in Fig.~\ref{fig:WFs_zeromode}~(c) are assigned to the  experimental $0_7^+$ state at 16.7 MeV and $0_8^+$ state at 18.8 MeV, respectively.
 The  reason why we  have the  two zero-mode $0^+$ excited states is because we treat the quantum fluctuations of  ${\hat Q}$ and  ${\hat P}$ properly and are naturally led to the zero-mode equation (\ref{eq:HuQPeigen}). The logic that more than two $0^+$ states appear at  low excitation energies from the Hoyle-analog state is shared by both the cases of
  $^{12}$C and $^{12}$O. This logic is   quite similar to the  case of a deformed nucleus, for which the rotational symmetry is spontaneously broken and the zero-mode states $2^+$, $4^+$, $6^+$, $\cdots$ are the members of the rotational band by increasing the  angular momentum.
In the case of the zero-mode excited states for the BEC, the quantum number counting the number of nodes in the $q$-space plays the role of the angular momentum.
Both of the emergences of the zero-mode excited states for the BEC of $\alpha$ clusters and the rotational band for the deformed nucleus are due to the finiteness of the systems. We note that in the infinite limit of
the system size of the BEC of $\alpha$ clusters, all the  zero-mode excited states become degenerate to zero energy.

%Fig5  different condensation rates energy levels
\begin{figure}[b]
\begin{center}
\includegraphics[width=8.6cm]{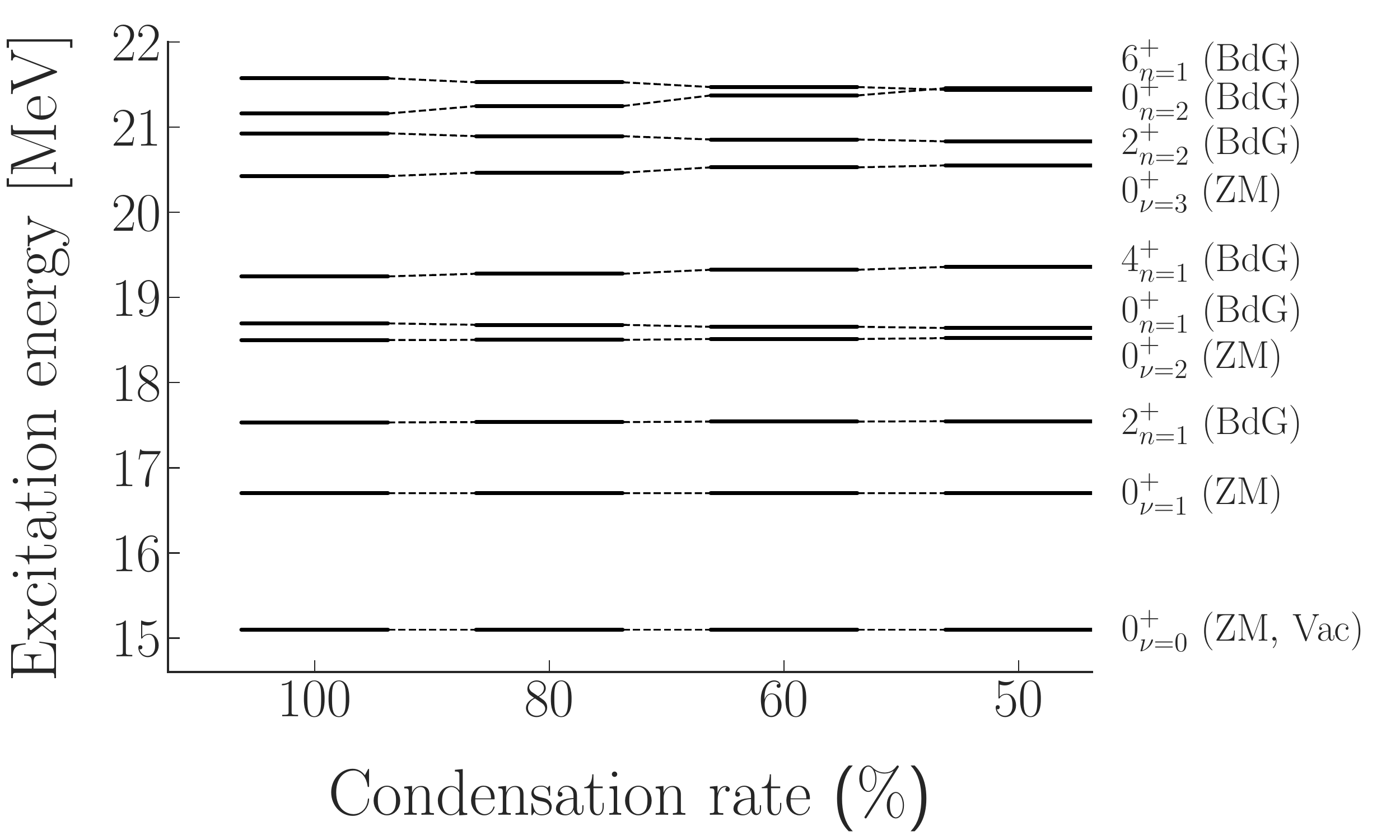}
\caption{The energy levels in $^{16}$O calculated
for different condensation rates and the experimental $\alpha$ cluster states.
}
\label{fig:Energy_levels_conden}
\end{center}
\end{figure}

 \par
 To see the  robustness of  the emergence of
  energy level structure  of  the  zero-mode and BdG states,  the calculated energy levels for different condensation rates,  50, 60, 80 and 100\%, are    displayed in Fig.~\ref{fig:Energy_levels_conden}.  While the parameter $\Omega$ is kept  to be 1.57 MeV$/ \hbar$, which was obtained for the case of 61\%, the parameter $V_r$ is determined so as to minimize the mean square errors between the experimental and calculated energies of the state $0_7^+$ and the states above it.
  The  energy level structure  is  little changed by the different condensation rates.  Especially the excitation energies of the two $0^+$ states above the vacuum
   are almost independent of the condensation rate as long as the BEC is formed.  Although the  condensation rate has not been known experimentally, it is clear that the level structure due to BEC in Fig.~\ref{fig:Energy_levels} is robust.

\par
In Table~\ref{table:Transition_Probabilities}, the   calculated $E0$ and $E2$ transition probabilities    are   listed.  Although no experimental data are available, the  transitions from the $2_1^+({\rm BdG})$ state to the $0_2^+$ ({\rm ZM}) and $0^+$ ({\rm Vac}) states are strong.
Experimental  measurements, which may serve to    check the condensation rate and the models, are desired.

 %%Table monopole BE2 Table I
 \begin{table}[t]
\caption{Calculated  reduced $E0$ and $E2$ transition probabilities, $B({\rm E}2)$
and   $M({\rm E}0:0^+\, \rightarrow\, 0^+)$
in $^{16}$O, in units of $e^2\,{\rm fm}^4$ and ${\rm fm}^2$, respectively, for 61$\%$  and 100$\%$ condensations.
}
\label{table:Transition_Probabilities}
	\begin{center}
		\begin{tabular}{l|r|r} \hline \hline
			Transition & 61\% & 100\% \\ \hline
			$M({\rm E}0:0^+_{\nu=1}({\rm ZM})\, \rightarrow \, 0^+_{\nu=0}({\rm ZM}))$ &4.2 &  7.6\\
  		$M({\rm E}0:0^+_{\nu=2}({\rm ZM})\, \rightarrow \, 0^+_{\nu=0}({\rm ZM}))$ &0.24 &  0.62\\
			$B({\rm E}2:2^+_{n=1}({\rm BdG})\, \rightarrow \, 0^+_{\nu=0}({\rm ZM}))$ &369 &606\\
			$B({\rm E}2:2^+_{n=1}({\rm BdG})\, \rightarrow \, 0^+_{\nu=1}({\rm ZM}))$ &1644 &  1724\\
%%%% added in the case of revision
%\textcolor{red}{
			$B({\rm E}2:4^+_{n=1}({\rm BdG})\, \rightarrow \, 2^+_{n=1}({\rm BdG}))$ &785 & 456\\
			$B({\rm E}2:6^+_{n=1}({\rm BdG})\, \rightarrow \, 4^+_{n=1}({\rm BdG}))$ &1111&  469\\
			$B({\rm E}2:8^+_{n=1}({\rm BdG})\, \rightarrow \, 6^+_{n=1}({\rm BdG}))$ &1355 &  448\\
%		$B({\rm E}2:10^+_{n=1}({\rm BdG})\, \rightarrow \, 8^+_{n=1})({\rm BdG})$ &1570 &1358\\
	 \hline\hline

\end{tabular}
	\end{center}
\end{table}

%Fig6  JJ+1 plot
\begin{figure}[t]
\begin{center}
\includegraphics[width=7.8cm]{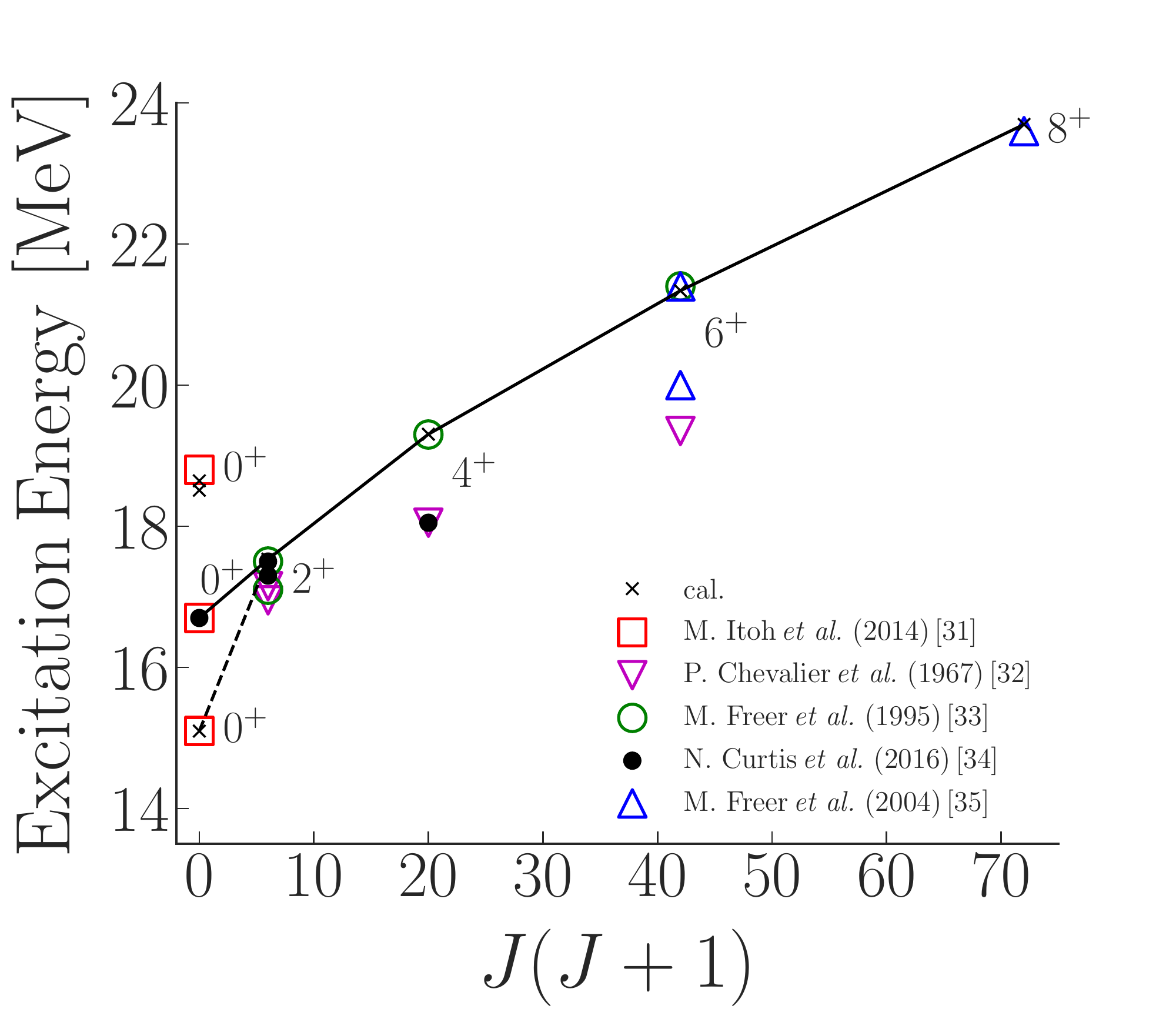}
\caption{(Color online) The calculated energy levels of the Nambu-Goldstone zero-mode states
and the BdG states are displayed against $J(J+1)$ in comparison with the experimental energy levels in $^{16}$O,
red square \cite{Itoh2014}, magenta down triangle \cite{Chevallier1967},  green open circle \cite{Freer1995},
black filled circle \cite{Curtis2016}, and blue up triangle \cite{Freer2004}.
The line is to guide the eye.
%for $(\Omega, V_r)=(1.57, 513)$ and $(1.60, 488)$ MeV
}
\label{fig:JJ+1}
\end{center}
\end{figure}
\par
%J(J+1) plot

In Fig.~6 the calculated energy levels of the Nambu-Goldstone zero-mode states and the BdG states are displayed against $J(J+1)$ in comparison with the experimental data. The calculated  BdG states $2^+$, $4^+$,  $6^+$and $8^+$are approximately located   on the straight line in  $J(J+1)$ plot with a rotational constant $k$=93 keV for the $2^+ \sim 8^+$ states, which seems to be  consistent with the values of $k=$95$\pm$ 20 from  the data  of Freer {\it et al.} \cite{Freer1995},  $k=$86 keV estimated  from each  centroid of the observed  $2^+$ , $4^+$ and $6^+$ states  in Ref.~\cite{Ohkubo2010} and   $k=$80 keV of the  calculated value
% by Ohkubo and Hirabayashi
  in Ref.~\cite{Ohkubo2010}
%, while  $k=$64 keV from the data of Chevallier {\it et al.}\cite{Chevallier1967},
 where $k $= $\hbar^2$/2$\mathcal{J}$  with  $\mathcal{J}$
being the moment of inertia.
This shows that the BdG states built on the Nambu-Goldstone state at 16.7 MeV are strongly deformed. The calculated large $B$(E2) values of the transitions in this band  in Table~I are  in accordance  with a strongly deformed BEC of $\alpha$ clusters.
As seen in Fig.~6,  it is also clear that the rotational band is built on the Nambu-Goldstone zero-mode  $0^+$ state at 16.7 MeV and not on the vacuum 15.1 MeV   $0^+$ state.
The logical structure of the emergence of a deformed BEC on the Nambu-Goldstone zero-mode  $0^+$ state  is essentially the same as that seen for the BEC of three $\alpha$ clusters, as has been suggested for $^{12}$C and $^{16}$O in Ref.~\cite{Ohkubo2013}.
In the picture of a local condensate with the $\alpha$+$^{12}$C(Hoyle state) structure in Ref.~\cite{Ohkubo2010}, the condensation rate is estimated to be 70\%$\times$12/16=53\%, which will be increased if the involvement  of the fourth $\alpha$ cluster to the condensation is taken into account. The present condensation rate around 60\% seems to be  consistent with this rough estimate.

\par
The above discussions and conclusions are reconfirmed  also by the other method of fitting the parameters, in which  the  rms radius of the Hoyle-analog state is constrained to  5 fm of  Ref.~\cite{Funaki2008C}, which gives the best values of  $\Omega$=1.67 MeV/$\hbar$ and $V_r=497$ MeV to fit the 16.7 MeV $0_7^+$ state.
We mention  that the present external  harmonic potential $V_{\rm ex}(r)$ with  $\Omega$ represents a phenomenological mean field potential of the  $\alpha$ clusters of Bose-Einstein condensate, which is accompanied by the residual ${\alpha-\alpha}$ interaction  $U$, and which corresponds to a ``container'' in  Ref.~\cite{Funaki2018} whose
 potential form is given neither microscopically nor phenomenologically.
By using a finite well such as a Woods-Saxon or a Woods-Saxon squared form factor for  $V_{\rm ex}(r)$,  the $\alpha$ decay width can be calculated. Therefore   the experimental determinations of the decay widths, especially  for the 16.7 MeV $0^+$ and 18.8 MeV $0^+$  states \cite{Itoh2014}, are highly desired.

 \par
 To summarize,
we have studied the observed well-developed $\alpha$ cluster states above the four $\alpha$ threshold from the viewpoint of  Bose-Einstein condensation of $\alpha$ clusters  using a field theoretical superfluid cluster model
in which  the order parameter is defined.
We could  reproduce the observed level structure of the $\alpha$ cluster states above the threshold  for the first time.  It is found that the emergence of the  level structure that the  two $0^+$ states with a well-developed $\alpha$ cluster structure at  very low excitation energies from the threshold is
a manifestation of the Nambu-Goldstone zero-mode states due to the BEC of the vacuum $0_6^+$  (15.1 MeV).
  This mechanism is the same to the $\alpha$ cluster structure above the three $\alpha$ threshold in $^{12}$C. It is found that the obtained level structure is robust and  changes little once the $\alpha$ cluster condensation is realized with a significant condensation rate. The present results give evidence of the existence of Bose-Einstein condensate of $\alpha$ cluster in $^{16}$O.

 \par
%\textit{Acknowledgements}
This work is supported by JSPS KAKENHI Grant No.~16K05488.
One of the authors (SO)  thanks the Yukawa Institute for Theoretical Physics at Kyoto University for   the hospitality extended  during  stays in  2017 and 2018.


\begin{thebibliography}{00}
\bibitem{Wefelmeier1937}
W. Wefelmeier, Z. Phys. {\bf 107}, 332 (1937).
\bibitem{Wheeler1937}
J.  A. Wheeler, Phys. Rev. {\bf  52}, 1083 (1937); Phys. Rev. {\bf  52}, 1107 (1937).
\bibitem{Ikeda1968} %Ikeda diagram 8Be-24Mg
K. Ikeda, N.~Takigawa, and H.~Horiuchi,
Prog. Theor. Phys. Suppl. {\bf E68}, 464 (1968).
\bibitem {Horiuchi1972} %Ikeda diagram  -32S
H. Horiuchi, K. Ikeda, and Y. Suzuki,
Prog. Theor. Phys. Suppl. {\bf 52}, 89 (1972).
\bibitem {Suppl1972}
K. Ikeda, T. Marumori, R. Tamagaki, and H. Tanaka,
Prog. Theor. Phys. Suppl. {\bf 52}, 1 (1972)
and references therein.
\bibitem {Wildermuth1977}
K. Wildermuth and Y. C. Tang,
{\it A Unified Theory of the Nucleus}
(Vieweg, Braunschweig, 1977).
\bibitem {Suppl1980}
K. Ikeda, H. Horiuchi, and S. Saito,
Prog. Theor. Phys. Suppl. {\bf 68}, 1 (1980)
and references therein.
\bibitem {Suppl1998}
S. Ohkubo, M. Fujiwara, and P. E. Hodgson,
Prog. Theor. Phys. Suppl. {\bf 132}, 1 (1998)
and references therein.
\bibitem {Ohkubo1989} %Ikeda diagram - 44Ti
S. Ohkubo, K. Umehara, and K. Hiraoka,
{\it Developments of nuclear cluster dynamics}
  (World Scientific, 1988) p.114.
\bibitem {Ohkubo1998} %Ikeda diagram - 56Ni
S. Ohkubo, T. Yamaya, and P. E. Hodgson,
Nuclear clusters, in {\it Nucleon-Hadron Many-Body
Systems}, (edited by H. Ejiri and H. Toki)
(Oxford University Press, Oxford,1999), p. 150.
\bibitem {Fujiwara1980}
Y.~Fujiwara {\it et al.},
Prog. Theor. Phys. Suppl. {\bf 68},  29 (1980).
\bibitem {Michel1998}
F. Michel, S. Ohkubo, and G. Reidemeister, Prog. Theor. Phys.
Suppl. {\bf 132}, 7 (1998).
\bibitem {Bohr1969B}
A. Bohr and B. R. Mottelson: {\it Nuclear Structure}, Vol. II,
(Benjamin, Inc., New York, 1975).
\bibitem {Ring1980}
P. Ring and P. Schuck,
{\it The nuclear many-body problem} (Springer-Verlag, Berlin, 1980).
\bibitem{Uegaki1977}
E. Uegaki, S. Okabe, Y. Abe, and H. Tanaka,
Prog.~Theor.~Phys. {\bf 57}, 1262 (1977);
E. Uegaki, Y. Abe, S. Okabe, and H. Tanaka, Prog.~Theor.~Phys. {\bf 62}, 1621 (1979).
\bibitem{Morinaga1956}
H. Morinaga,
Phys. Rev. {\bf 101}, 254 (1956).
\bibitem {Matsumura2004}
H.  Matsumura and Y. Suzuki,
Nucl. Phys. {\bf A 739}, 238 (2004).
\bibitem {Kurokawa2004}
C. Kurokawa and K. Kato, Nucl. Phys. {\bf A 738}, 455 (2004).
\bibitem {Yamada2004}
T. Yamada and P. Schuck, Phys. Rev. C {\bf 69}, 024309 (2004).
\bibitem {Ohtsubo2013}
S. Ohtsubo, Y. Fukushima, M. Kamimura, and E. Hiyama,
Prog. Theor. Exp. Phys. 073D02 (2013).
\bibitem {Tohsaki2001}
A. Tohsaki, H. Horiuchi, P. Schuck, and G. R$\ddot{\rm o}$pke,
Phys.~Rev.~Lett. {\bf 87}, 192501 (2001).
%ab initio 12C Hoyle
\bibitem {Kanada2007} % 12C
Y. Kanada-En'yo,  Prog. Theor.  Phys. {\bf 117}, 655 (2007).
\bibitem {Chernykh2007}
M. Chernykh, H. Feldmeier, T. Neff, P. von Neumann-Cosel, and A. Richter,
Phys. Rev. Lett. {\bf 98}, 032501 (2007).
\bibitem {Roth2011}%Structure and Rotations of the Hoyle State Energy
R. Roth, J. Langhammer, A. Calci, S. Binder, and P. Navratil,
 Phys. Rev. Lett. {\bf 107}, 072501 (2011).
\bibitem {Dreyfuss2013}%sipletic model  Rotations of the Hoyle State Evgeny
A. C. Dreyfuss, K. D. Launey, T. Dytrych, J. P. Draayer, and C. Bahri,
Phys. Lett. {\bf B 727}, 515 (2013).
\bibitem {Epelbaum2012}%Structure and Rotations of the Hoyle State Evgeny
E. Epelbaum, H. Krebs, T. A. Lahde, D. Lee, and Ulf-G. Meissner,
Phys. Rev. Lett. {\bf 109}, 252501 (2012).
\bibitem{Funaki2015} %Hoyle band 12C
Y. Funaki, Phys. Rev. C {\bf 92}, 021302 (R) (2015).
% 4 alpha 16O N=4
\bibitem {Funaki2016} % 12C Hoyle monopole
Y. Funaki,
Phys. Rev. C {\bf 94}, 024344 (2016).
\bibitem {Nakamura2016}
Y.~Nakamura, J.~Takahashi, Y.~Yamanaka, and S.~Ohkubo,
Phys.~Rev.~C {\bf 94}, 014314 (2016); {\bf 98}, 049901(E) (2018).
\bibitem {Katsuragi2018}
R.~Katsuragi, K.~Kazama, J.~Takahashi, Y.~Nakamura, Y.~Yamanaka, and S.~Ohkubo,
Phys.~Rev.~C {\bf 98}, 044303 (2018) and earlier references therein.
\bibitem{Itoh2014} %16O 4 alpha search
M. Itoh {\it et al.},
J. Phys. Conf. Ser. {\bf 569}, 012009 (2014).
\bibitem{Chevallier1967} % 4 alpha chain
P.~Chevallier, %{\it et al.},
 F.~Scheibling, G.~Goldring, I.~Plesser, and M.~W.~Sachs,
Phys. Rev. {\bf 160}, 827 (1967).
\bibitem{Freer1995} %16O 4 alpha search 12C(16O,4alpha)
M. Freer  {\it et al.},
%N. M. Clarke, N. Curtis, B. R. Fulton, S. J. Hall, M. J. Leddy, J. S. Pople, G. Tungate, R. P. Ward, P. M. Simmons, W. D. M. Rae, S. P. G. Chappell, S. P. Fox, C. D. Jones, D. L. Watson, G. J. Gyapong, S. M. Singer, W. N. Catford, and P. H. Regan,
Phys. Rev  C {\bf 51}, 1682 (1995).
\bibitem{Curtis2016} %16O 4 alpha search
N. Curtis {\it et al.},
%S. Almaraz-Calderon, A. Aprahamian, N. I. Ashwood, M. Barr, B. Bucher, P. Copp, M. Couder, X. Fang, M. Freer, G. Goldring, F. Jung, S. R. Lesher, W. Lu, J. D. Malcolm, A. Roberts, W. P. Tan, C. Wheldon, and V. A. Ziman,
Phys. Rev  C {\bf 94}, 034313 (2016).
\bibitem{Freer2004}% 8Be+8Be decay of 16O
M. Freer  {\it et al.},
%M. P. Nicoli, and S. M. Singer, C. A. Bremner, S. P. G. Chappell, W. D. M. Rae,  I. Boztosun, B. R. Fulton, D. L. Watson, B. J. Greenhalgh, G. K. Dillon,  R. L. Cowin, and D. C. Weisser,
Phys. Rev  C {\bf 70}, 064311 (2004).
\bibitem{Freer2005} %alpha particle states in 16O and 20Ne short contribution
M. Freer, % for charrisa collaboration
J. Phys. {\bf G31},  S1795 (2005).
\bibitem{Funaki2008C}
Y.~Funaki, T.~Yamada, H.~Horiuchi, G.~R\"{o}pke, P.~Schuck, and A.~Tohsaki,
Phys. Rev. Lett. {\bf 101}, 082502 (2008).
\bibitem{Ohkubo2010}
S. Ohkubo and Y. Hirabayashi,
Phys. Lett. {\bf B 684}, 127 (2010).
\bibitem {Funaki2018} %16O container
Y. Funaki,
Phys. Rev. C {\bf 97},  021304(R) (2018).
\bibitem {ENSDF}
D. R. Tilley, H. R. Weller, and C. M. Cheves,
Nucl. Phys. {\bf A565}, 1 (1993);
Brookhaven National Nuclear Data Center, http://www.nndc.bnl.gov/ensdf/.
% 4 alpha linear chain
\bibitem{Suzuki1972}
Y.~Suzuki, H.~Horiuchi, and K.~Ikeda,
Prog.~Theor.~Phys. {\bf 47}, 1517 (1972).
\bibitem{Ichikawa2011}
T. Ichikawa, J. A. Maruhn, N. Itagaki, and S. Ohkubo
Phys. Rev. Lett. {\bf 107}, 112501 (2011).
\bibitem{Horiuchi2017}
W. Horiuchi and Y.~Suzuki,
Phys. Rev. C {\bf 95}, 044320 (2017).
\bibitem{Inakura2018}
T. Inakura and S. Mizutori,
Phys. Rev  C {\bf 98}, 044312 (2018).
\bibitem{Curtis2013} %4 alpha linear chain search  failed
N. Curtis  {\it et al.},
%S. Almaraz-Calderon, A. Aprahamian, N. I. Ashwood, M. Barr, B. Bucher, P. Copp, M. Couder, X. Fang, M. Freer, G. Goldring, F. Jung, S. R. Lesher, W. Lu, J. D. Malcolm, A. Roberts, W. P. Tan, C. Wheldon, and V. A. Ziman,
Phys. Rev. {\bf C} 88, 064309 (2013).
%
\bibitem {Ohkubo2013}
S. Ohkubo, arXiv: nucl-th 1301.7485 (2013).
\bibitem {Suzuki1976} %OCM
Y.~Suzuki,
Prog. Theor. Phys. {\bf 55}, 1751 (1976);
 Prog. Theor. Phys. {\bf 56}, 111 (1976).
\bibitem {Ali1966}
S. Ali and A. R. Bodmer,
Nucl.~Phys. {\bf A 80}, 99 (1966).
\end{thebibliography}
\end{document}